\documentclass[conference]{IEEEtran}
\IEEEoverridecommandlockouts
\usepackage{cite}
\usepackage{amsmath,amssymb,amsfonts}
\usepackage{algorithmic}
\usepackage{graphicx}
\usepackage{textcomp}
\usepackage{xcolor}

\usepackage{balance}
\usepackage{flushend}
\usepackage{csquotes}
\usepackage{hyperref}
\usepackage{xcolor}
\usepackage{xurl}

\hypersetup{
    colorlinks,
    linkcolor={black!50!black},
    citecolor={black!50!black},
    urlcolor={black!80!black}
}

\def\BibTeX{{\rm B\kern-.05em{\sc i\kern-.025em b}\kern-.08em
    T\kern-.1667em\lower.7ex\hbox{E}\kern-.125emX}}
\begin{document}

\title{Terminal Lucidity:\\Envisioning the Future of the Terminal
}

\author{
\IEEEauthorblockN{Michael MacInnis}
\IEEEauthorblockA{School of Computer Science\\
Carleton University, Ottawa, Canada\\
}
\and
\IEEEauthorblockN{Olga Baysal}
\IEEEauthorblockA{School of Computer Science\\
Carleton University, Ottawa, Canada\\
}
\and
\IEEEauthorblockN{Michele Lanza}
\IEEEauthorblockA{{REVEAL @ Software Institute}\\
\textit{USI, Lugano, Switzerland}\\
}
}

\maketitle

\begin{abstract}

The Unix terminal, or just simply, the terminal, can be found being applied in almost every facet of computing. It is available across all major platforms and often integrated into other applications. Due to its ubiquity, even marginal improvements to the terminal have the potential to make massive improvements to productivity on a global scale. We believe that evolutionary improvements to the terminal, in its current incarnation as windowed terminal emulator, are possible and that developing a thorough understanding of issues that current terminal users face is fundamental to knowing how the terminal should evolve.
In order to develop that understanding we have mined Unix and Linux Stack Exchange using a fully-reproducible method which was able to extract and categorize 91.0\% of 1,489 terminal-related questions (from the full set of nearly 240,000 questions) without manual intervention.

We present an analysis, to our knowledge the first of its kind, of windowed terminal-related questions posted over a 15-year period and viewed, in aggregate, approximately 40 million times. As expected, given its longevity, we find the terminal's many features being applied across a wide variety of use cases. We find evidence that the terminal, as windowed terminal emulator, has neither fully adapted to its now current graphical environment nor completely untangled itself from features more suited to incarnations in previous environments. We also find evidence of areas where we believe the terminal could be extended along with other areas where it could be simplified. Surprisingly, while many current efforts to improve the terminal include improving the terminal's social and collaborative aspects, we find little evidence of this as a prominent pain point.






\end{abstract}


\section{Introduction}

From humble beginnings as a teletypewriter, through video display terminals, to windowed terminal emulators, what we now think of as the Unix terminal, or just simply, the terminal, has become a fixture of the modern computing environment. The terminal is everywhere.

Before Windows Subsystem for Linux (WSL), the Microsoft POSIX subsystem, Cygwin, and Windows Services for UNIX provided a stopgap for those requiring access to Unix command-line tools on Windows~\cite{Wikipedia:WindowsServicesForUNIX}. WSL might have provided a similar anemic experience if not for the Windows Terminal 
which brought a full-featured, integrated, Unix terminal to Windows.\footnote{\url{https://www.youtube.com/watch?v=8gw0rXPMMPE}}. With the release of Windows 11, the Windows Terminal is now the default command-line environment on Windows~\cite{WindowsTerminalDefaultInWindows11} making the terminal available, by default, across all major platforms.

The terminal is used so frequently when working with containers and the cloud that introductions to Docker or Kubernetes often contain inline discussions of the Unix shell and how to run commands from the terminal~\cite{DockerInAction2ndEdition, KubernetesInAction}. While Python dominates the AI space (with C++ under the hood) and large language model (LLM) experimentation often assumes the presence of Jupyter Notebooks, the terminal is used to perform tasks that fall outside the purview of these tools~\cite{HowToCreateACustomLanguageModel}. Many modern text editors, such as Zed, feature an integrated terminal~\cite{Zed}. Installation instructions for languages like Go and Rust assume the presence of a terminal~\cite{GoDownloadAndInstallation,RustDownloadAndInstallation}. The Integrated Development Environment (IDE), once the antithesis of the terminal-based approach, now often provides an integrated terminal. Microsoft's VS~Code IDE includes an integrated terminal~\cite{IntegratedTerminalInVSCode}. KDevelop includes an integrated terminal~\cite{Konsole}. JetBrain's IDEs \& Android Studio include an integrated terminal~\cite{GetToKnowTheAndroidStudioUI, NewTerminalJetBrains} and the Android SDK includes a suite of command-line tools~\cite{AndroidStudioCommandLineTools}.

There are several command-line iOS development tools, and while Xcode does not currently provide an integrated terminal, iOS applications are developed on macOS which, due to its UNIX heritage, provides a terminal. Even before OS X the terminal was a prominent part of \enquote{the premier software development system for the Mac}~\cite{InTheBeginning} and when Mac OS X was introduced, the Macintosh became a \enquote{UNIX box}, and the terminal was front and center in early promotional material.\footnote{\url{https://www.brainmapping.org/MarkCohen/UNIXad.pdf}}

We do not wish to overstate the importance of the terminal. We are not claiming anything close to \enquote{real developers use the terminal}. Using the terminal where it does not make sense is as strange as avoiding it where it does. We do, however, find that, given its ubiquity and utility in areas like software and systems development~\cite{TheGoProgrammingLanguage, DockerInAction2ndEdition, KubernetesInAction, NetworkProgrammabilityAndAutomation}, system administration~\cite{UNIXAndLinuxSystemAdministration}, performance tuning~\cite{SystemsPerformance}, data science~\cite{DataScienceAtTheCommandLine}, computer and network security~\cite{LearningDevSecOps}, etc. the terminal is often conspicuously absent from research discussions in these areas. We can only speculate as to why. It might be a case of familiarity breeding contempt. It could be that the terminal has become so common as to be invisible. We reject the idea that the terminal has achieved perfection with no potential for improvement.

The terminal deserves credit for its longevity and continued relevance over the last 50 years but it has not existed unchanged. The terminal is not even one entity but rather a large set of distinct applications with features that have converged to support popular terminal-based applications. We believe further evolution of the terminal is possible and that the key to understanding how the terminal should evolve lies in understanding the issues that current terminal users face.

The primary question we wish to address with this work is the possibility of developing a requirements-driven understanding of how the terminal should evolve based on a systematic analysis of a large body of windowed terminal-related questions posted over a 15-year period and viewed, in aggregate, approximately 40 million times.



\section{Related Work} \label{related-work}

While there are many papers that mine the Stack Exchange network for insights, such as \cite{StackExchangePaper1, StackExchangePaper2, StackExchangePaper3, StackExchangePaper4}, we know of no work that is directly related to the topic of this paper --- investigating terminal-related issues to understand how the terminal should evolve. Other than~\cite{TerminalsAllTheWayDown}, we know of no recent terminal-related papers. We feel that this absence combined with the popularity and longevity of the terminal, only serves to underscore the urgent need for more work in this area.

Popular terminals tend to be fairly conservative in the features they offer. There are, however, at any given time, often several efforts, at various stages in development, attempting to pull the terminal in new directions. Recent efforts include Xiki, Fig, sshx, and Warp. Xiki, a \enquote{shell console with GUI features}, promises a friendlier and more powerful command line~\cite{GitHub:Xiki}, but development seems to be stalled. Fig, the \enquote{next-generation command line}, aimed to make \enquote{the command line easier and more collaborative}~\cite{FigManual}. Unfortunately, Fig is sunsetting effective September 1st, 2024~\cite{Fig}. Sshx, an application that lets users \enquote{share collaborative terminals over the web, with live cursors on an infinite canvas}~\cite{Reddit:sshx} is being actively developed~\cite{Github:sshx}. Warp, a \enquote{terminal built for the 21st century}, is currently in public beta~\cite{Warp}. A goal many of these efforts share appears to be improving the social and collaborative aspects of the terminal.

\section{Methodology}

To understand the issues that terminal users face we analyzed terminal-related questions extracted from Unix \& Linux Stack Exchange (U\&LSE).\footnote{\url{https://unix.stackexchange.com/}} As of March 2023, U\&LSE was one of \enquote{the three most actively-viewed sites} in the Stack Exchange network~\cite{Wikipedia:StackExchange}. Topics discussed on U\&LSE overlap with those discussed on Ask Ubuntu, Server Fault, Super User, and Stack Overflow but whereas those sites focus on a particular Linux distribution, system administrators, power users, and programmers, respectively, questions on U\&LSE are of interest to an intersection of those audiences. There is no dataset that fits our purposes exactly. U\&LSE works as a rough approximation and is more suited than other similar datasets.

\subsection{Selection and Filtering}

Browsing the April 2024 U\&LSE Data Dump, we observed that all of the potentially interesting questions we manually identified contained the word (or tag) \emph{terminal}. We initially resisted filtering out closed questions and only performed minimal filtering based on score as we did not want to eliminate questions that might suggest more radical departures from the state of the art. What we found was that more radical questions often had fairly high scores. Low-scoring and/or closed questions, that were not completely unrelated, were often unclear and sometimes poorly-worded, near-duplicates of questions with higher scores. In line with other investigations~\cite{QuestionScore:2016}, we found that eliminating closed questions significantly reduced noise and that a question score threshold could be used as a \enquote{slider} to reduce not only the noise but the size of the filtered data set.

Restricting our analysis to questions containing the word (or tag) \emph{terminal}, or, for good measure, the name of any one of 20 popular terminal emulators (\emph{iterm}, \emph{konsole}, \emph{rxvt}, \emph{xterm}, etc., see the replication package for the full list) --- reduced the set of 238,249 questions to 20,805 questions. Due to pragmatic considerations, further reductions of the data set were necessary. We initially considered random sampling but ultimately decided against it so as not to hinder the reproducibility of our results. Working backwards, we found that questions with a score of less than ten provided \enquote{no new or relevant data}~\cite{BasicsOfQualitativeResearch}. Excluding closed questions and questions with a single-digit score or less resulted in a set of 1,489 potentially interesting questions.




\subsection{Discovering Local Regularity}

Browsing this smaller set of questions, common tags, words, and phrases became more apparent. The language used by members of a community develops \enquote{a high level of local regularity}~\cite{NaturalnessOfSofware}. The U\&LSE community has developed its own local dialect. The U\&LSE community is also a subset and reflection of the wider (and much older) Unix and Linux community which has an already established jargon. In addition to common words and phrases, the U\&LSE community has been developing a \enquote{folksonomy}~\cite{Wikipedia:Folksonomy} by applying tags to posts over the last 15 years.


\subsection{Computer-Aided Qualitative Coding}

Inspired by recent regex-based~\cite{InteractiveRegexQualitativeCoding} and hybrid~\cite{IntelligentCoRaters} qualitative coding efforts in quantitative ethnography and similar experiences in software engineering research~\cite{LinkingEmailsAndSourceCode:2009,RegexStackOverflow:2020} we created simple patterns to extract and sort questions into categories. Where possible we preferred to base our patterns on existing tags so as to reuse Stack Exchange’s folksonomy. Next we based our patterns on words and short phrases that appeared in multiple related posts. Where the phrasing of a post was too idiosyncratic to be generalized we based our pattern on the exact phrase that we believed identified a post as being related to a category. Our automated process then sorted the data into their respective categories and the posts in those categories were manually reviewed. 

Patterns were adjusted (which sometimes involved adding or removing categories) and the scripts run again. By using \verb|git| (and because Stack Exchange posts are single-line records), we were easily able to see the effect of these changes across hundreds of iterations as we verified the categories we discovered \enquote{over and over again against actual data}~\cite{BasicsOfQualitativeResearch}. Guiding each iteration was the goal of using the fewest number and smallest patterns possible while minimizing the number of manual corrections required - minimizing the “disagreements” between the human and the automated process.

Computer-aided qualitative coding allows one “rater” to compare and repeatedly refine their results against an automated process. In our case, the result of this refinement is a set of simple regular expressions able to sort 1,355 of 1,489 questions into our 26 discovered categories without manual intervention for an inter-rater reliability of 91\%. By restricting ourselves to regular expressions we are able to present clear and unambiguous rules. While the patterns we created provide transparency they are overly simplistic in the face of unrestricted natural language. The categorization of 83 questions (5.6\%) had to be manually corrected. A further 51 questions (3.4\%) were assigned a correct category along with one or more incorrect categories and had to be manually adjusted. Manually adjusted questions are marked with an asterisk ($^*$) when referenced in the results section. Manually corrected categorizations are detailed in the Appendix.

\begin{figure}[ht]
    \centering
    \includegraphics[width=0.9\columnwidth]{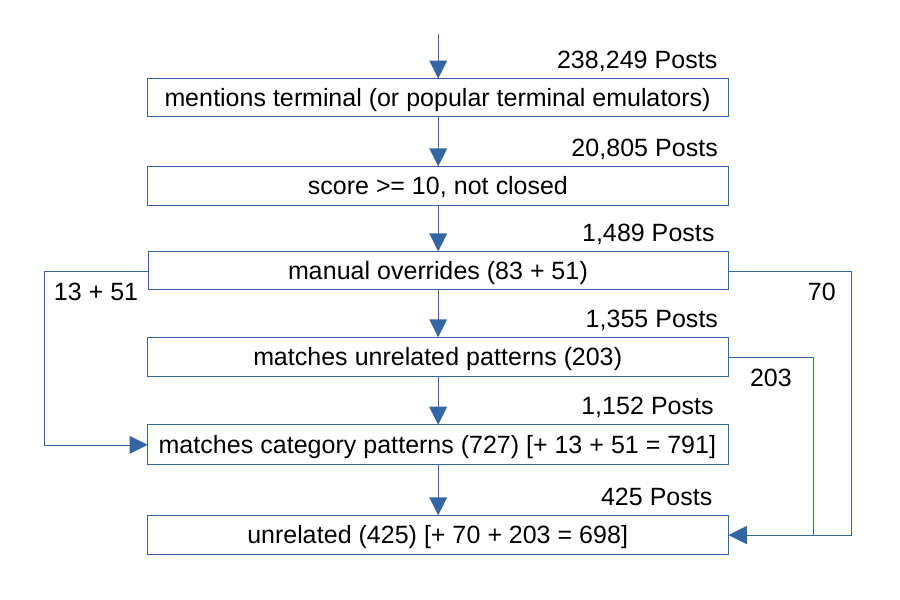}
    \caption{Question categorization process.}
    \label{fig:QuestionCategorizationProcess}
\end{figure}

Other than the manually categorized questions, which are extracted before further processing there are no ordering restrictions on the patterns for categories other than the unrelated category. The unrelated category is a combination of questions that match a set of patterns as well as the unselected questions that fall through the pipeline. The set of patterns for the unrelated category is reliable enough that we use them as a pre-filter. The categorization process is depicted in Figure~\ref{fig:QuestionCategorizationProcess}.

\subsection{Caveats}

While we have made every effort to ensure the consistency of our categorization we make no claims regarding its applicability beyond the discussion in this paper or datasets other than U\&LSE. 

The primary advantage of our approach is that it is possible for it to be \enquote{independently replicated by other researchers}\cite{CommonMethodologicalMistakes}.
All of the scripts and patterns used in our analysis are available in the replication package along with instructions on their use.\footnote{\url{https://doi.org/10.5281/zenodo.12788678}}

\section{Results} \label{results}

\begin{figure}[ht]
    \centering
    \includegraphics[width=\columnwidth]{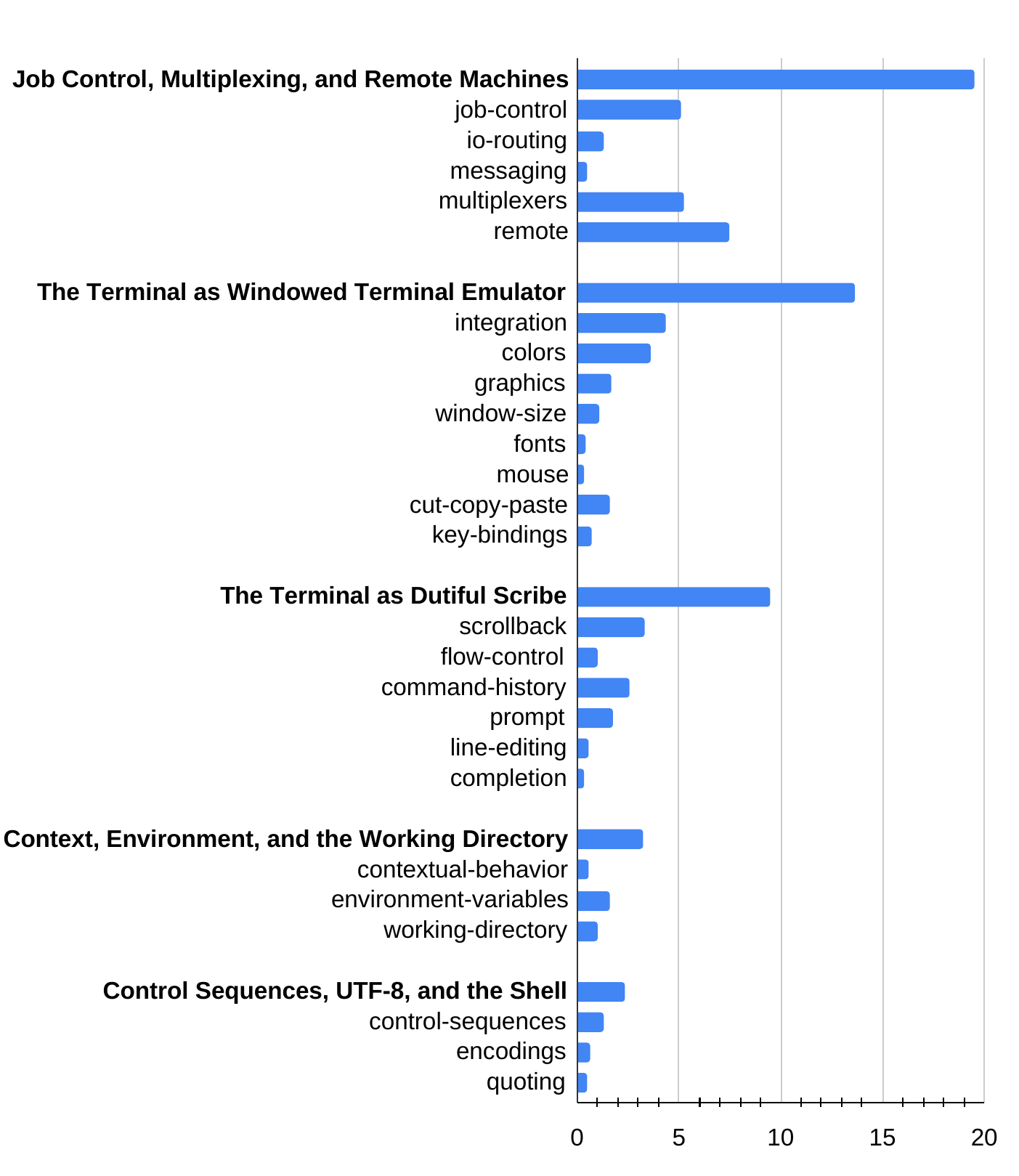}
    \caption{Question views (in millions) per category.}
    \label{fig:ViewsPerCategory}
\end{figure}

The following is a summary of 1,489 terminal-related questions belonging to 25 of our 26 discovered categories. The only category not shown is the unrelated category. We have grouped and ordered categories in an effort to make the information more easily digestible (see Figure \ref{fig:ViewsPerCategory}). Category names are given in each section along with the number of posts and aggregate number of views. 104 posts were assigned two categories, 11 posts were assigned three categories, and two posts were assigned four categories resulting in an inflated total of 923 posts for the 25 categories shown below and not 791 which is the number of unique posts not categorized as unrelated.

\subsection{Job Control, Multiplexing, and Remote Machines}

The most frequently viewed, terminal-related, posts on U\&LSE are related to job control, terminal multiplexers, and using the terminal to interact with remote systems.


\subsubsection{\textbf{job-control (posts: 76, views: 5,061,222)}}

First added to the 4BSD C shell in concert with 4BSD's new terminal driver, job control transformed the terminal into a multitasking interface.

\begin{quote}
Some people claim that a well-designed windowing system removes any need for job control. Some complain that the implementation of job control — requiring support from the kernel, the terminal driver, the shell, and some applications — is a hack. Some use job control with a windowing system, claiming a need for both~\cite{APUE3}.
\end{quote}

\smallskip

The claim that \enquote{a well-designed windowing system removes any need for job control} is an exaggeration, though job control is less useful in the presence of a windowing system, with some users looking to \enquote{disable job control entirely} (\href{https://unix.stackexchange.com/questions/137915}{137915}).
A major problem with job control and managing processes from the terminal in general, is that the mechanisms involved (controlling terminal, process groups, signals) are not widely understood, are under-specified, interact in complex and surprising ways, and allow applications to both override behavior and perform the same operations themselves. The shell itself complicates the situation by defaulting to delivering signals like \verb|SIGINT| to the currently running process which can make tasks like \enquote{[t]erminating an infinite loop} (\href{https://unix.stackexchange.com/questions/42287}{42287}) more difficult than one might expect. All of this leaves users wondering how to perform relatively simple tasks like running \enquote{a command which will survive terminal close} (\href{https://unix.stackexchange.com/questions/4004}{4004}) or how to \enquote{detach a process from a bash script} (\href{https://unix.stackexchange.com/questions/269805}{269805}). While it is possible to \enquote{detach} a process from the terminal, the reverse is not easily or reliably possible (\href{https://unix.stackexchange.com/questions/4034}{4034}). There is the question of what to do with output from a program that is no longer connected to a terminal (\href{https://unix.stackexchange.com/questions/17648}{17648}), or worse, a program not connected to a terminal that attempts to read input (\href{https://unix.stackexchange.com/questions/25024}{25024}). The terminal's buffering and the speed at which commands can generate output combine to make job control features seem less effective (\href{https://unix.stackexchange.com/questions/176917}{176917}).


a) \textbf{io-routing (posts: 7, views: 1,269,311)}
Just on the edge of what is possible, there is a desire to \enquote{attach a terminal to a detached process} (\href{https://unix.stackexchange.com/questions/31824}{31824}) without the use of a multiplexer, \enquote{view the output of a running process} (\href{https://unix.stackexchange.com/questions/58550}{58550}) from another terminal, and interact with processes in other terminals (\href{https://unix.stackexchange.com/questions/365256}{365256}).


b) \textbf{messaging (posts: 5, views: 462,386)}
Overlapping to a certain degree with the desire to interact with processes in other terminals, the terminal provides ancient communication facilities used by the \verb|mesg|, \verb|wall|, and \verb|write| commands. With the exception of \verb|wall|, these commands are now rarely used and often disabled on the receiving end (\href{https://unix.stackexchange.com/questions/135108}{135108}). This along with narrow scope of these tools typically leads to a search for alternatives (\href{https://unix.stackexchange.com/questions/21334}{21334}, \href{https://unix.stackexchange.com/questions/99460}{99460}, \href{https://unix.stackexchange.com/questions/377768}{377768}).


\subsubsection{\textbf{multiplexers (posts: 107, views: 5,229,577)}}

Terminal multiplexers like GNU screen~\cite{GNUScreen} and tmux~\cite{GitHub:tmux} enable launching multiple virtual terminals in one terminal session as well as detaching and reattaching to a session. They also allow users to \enquote{split the terminal into more than one 'view'} (\href{https://unix.stackexchange.com/questions/7453}{7453}), group terminals into different sessions (\href{https://unix.stackexchange.com/questions/24274}{24274}), share \enquote{a terminal with multiple users} (\href{https://unix.stackexchange.com/questions/122801}{122801}), programmatically send input to a terminal (\href{https://unix.stackexchange.com/questions/5305}{5305}), and make a terminal read-only (\href{https://unix.stackexchange.com/questions/7588}{7588}).
Not all users are \enquote{sold} on the concept of terminal multiplexers with some asking questions like, \enquote{Is screen useful?} (\href{https://unix.stackexchange.com/questions/4116}{4116}). The author of the kitty terminal has stated that \enquote{multiplexers add unnecessary overhead [and] suffer from a complexity cascade}~\cite{Github:KovidGoyalOnTerminalMultiplexers}. Terminal multiplexers do add another layer which adds complexity and the features multiplexers provide sometimes overlap with those provided by other layers which can leave users wondering how to \enquote{make Screen scroll like a normal terminal} (\href{https://unix.stackexchange.com/questions/43229}{43229}) or how to copy and \enquote{paste text selections between tmux and the clipboard} (\href{https://unix.stackexchange.com/questions/67673}{67673}). Terminal multiplexers often require some tweaking (\href{https://unix.stackexchange.com/questions/29907}{29907$^*$}, \href{https://unix.stackexchange.com/questions/38402}{38402}, \href{https://unix.stackexchange.com/questions/38911}{38911}, \href{https://unix.stackexchange.com/questions/348913}{348913}, \href{https://unix.stackexchange.com/questions/511267}{511267}). How to modify the default keybindings (\href{https://unix.stackexchange.com/questions/5998}{5998}, \href{https://unix.stackexchange.com/questions/453466}{453466}) to be less intrusive (\href{https://unix.stackexchange.com/questions/1636}{1636}) or to avoid conflicts with other applications (\href{https://unix.stackexchange.com/questions/12455}{12455}) are frequent questions. Given their complexity, it is not always obvious which operations are required to achieve a desired effect or worse if the desired effect is possible at all. However, even with their additional complexity, terminal multiplexers are undeniably useful, particularly in a remote context. 


\subsubsection{\textbf{remote (posts: 66, views: 7,499,998)}}

The terminal has always been a portal. Many applications and protocols have been developed over the years to extend the terminal's ability to communicate with a remote machine with SSH (and SCP and SFTP) now dominating. Within the confines of the terminal emulator window, SSH provides the illusion of interacting with a remote machine as if it were the local machine. This illusion breaks down when operations require access to both local and remote file systems with a common question being, \enquote{How to copy a file from a remote server to a local machine?} (\href{https://unix.stackexchange.com/questions/188285}{188285}) and vice-versa. The situation is a continuation of the Telnet/FTP divide (\href{https://unix.stackexchange.com/questions/35538}{35538}). While SSH can be used to make multiple \enquote{hops} from one remote machine to another, it is largely up to the user to remember how many \enquote{levels} there are in the current session if they wish \enquote{to exit everything except the 'root' shell} (\href{https://unix.stackexchange.com/questions/3212}{3212}) --- this problem is not unique to SSH but is more apparent in a remote context. SSH can also be used to forward X connections, allowing remote access to graphical applications (\href{https://unix.stackexchange.com/questions/7226}{7226}) although not without issues (\href{https://unix.stackexchange.com/questions/505908}{505908}) and with no completely backward-compatible Wayland equivalent (\href{https://unix.stackexchange.com/questions/776767}{776767}). 

SSH itself can be the cause of other issues, like the terminal appearing to hang when a connection is lost (\href{https://unix.stackexchange.com/questions/196701}{196701}) or interactive terminal-based programs not working correctly because no pseudo-terminal has been allocated (\href{https://unix.stackexchange.com/questions/269474}{269474}). SSH provides its own solutions to these issues, with mechanisms to break away from \enquote{an unresponsive ssh session} (\href{https://unix.stackexchange.com/questions/2919}{2919}) and force pseudo-terminal allocation (\href{https://unix.stackexchange.com/questions/43496}{43496}). It is somewhat remarkable that many terminal features, such as setting the window title (\href{https://unix.stackexchange.com/questions/14113}{14113}) or changing terminal colors, continue to work as expected over SSH connections. However, this does mean that remote systems may set the window title (\href{https://unix.stackexchange.com/questions/40830}{40830}) or change terminal colors (\href{https://unix.stackexchange.com/questions/58982}{58982}). There is also the question of how agnostic SSH should be (\href{https://unix.stackexchange.com/questions/205567}{205567}). 

\subsection{The Terminal as Windowed Terminal Emulator}

The windowed terminal emulator brought the terminal into the modern computing environment and the world of the graphical user interface.


\subsubsection{\textbf{integration (posts: 93, views: 4,321,881)}}

Placing the terminal in a graphical environment alongside other applications creates an expectation that the terminal should integrate and be able to interact with both the environment and other applications. There is a desire to seamlessly \enquote{launch GUI applications from a terminal window} (\href{https://unix.stackexchange.com/questions/150706}{150706}) and similarly to close \enquote{various windows using the terminal} (\href{https://unix.stackexchange.com/questions/85205}{85205}). Users wish to configure not only the terminal itself \enquote{from within the terminal} (\href{https://unix.stackexchange.com/questions/176175}{176175}) but, in general, to be able to perform tasks that can done outside the terminal from within the terminal. For example, being able to \enquote{set the audio volume using the terminal} (\href{https://unix.stackexchange.com/questions/32206}{32206}), \enquote{change the icon of an application's window} (\href{https://unix.stackexchange.com/questions/179174}{179174}), or \enquote{[c]hange desktop wallpaper} (\href{https://unix.stackexchange.com/questions/59653}{59653}) from the terminal. Many terminals now support \enquote{clickable links} (\href{https://unix.stackexchange.com/questions/112267}{112267}) and while it is possible to make file names clickable (\href{https://unix.stackexchange.com/questions/63417}{63417}) this is not the default behavior. 

Being a graphical application itself, common questions include how to \enquote{set the default size and position} (\href{https://unix.stackexchange.com/questions/48984}{48984}) for a new terminal window, configure a \enquote{shortcut to launch a terminal window} (\href{https://unix.stackexchange.com/questions/47608}{47608}), \enquote{launch an application with [the] default 'terminal emulator'} (\href{https://unix.stackexchange.com/questions/32547}{32547}), and configure the \enquote{Open in Terminal} feature of other applications (\href{https://unix.stackexchange.com/questions/336368}{336368}, \href{https://unix.stackexchange.com/questions/547492}{547492}). There are many looking for ways to launch a set of programs in multiple terminal windows or tabs (\href{https://unix.stackexchange.com/questions/22682}{22682}, \href{https://unix.stackexchange.com/questions/75902}{75902}, \href{https://unix.stackexchange.com/questions/158434}{158434}, \href{https://unix.stackexchange.com/questions/492365}{492365}). With the amount of configuration and customization possible a common question is not only how to save the configuration of a single terminal (\href{https://unix.stackexchange.com/questions/35748}{35748}) but the configuration (including the size and placement) for a set of windows (\href{https://unix.stackexchange.com/questions/76497}{76497}, \href{https://unix.stackexchange.com/questions/274534}{274534}, \href{https://unix.stackexchange.com/questions/593778}{593778}).


\subsubsection{\textbf{colors (posts: 63, views: 3,569,541)}}

The terminal, as video display terminal, developed techniques for displaying color. Some of these survive in, and have subsequently been extended by, windowed terminal emulators. Even with widespread support for color, it is difficult to reliably detect the level of support, such as whether the \enquote{terminal supports 24-bit / true color} (\href{https://unix.stackexchange.com/questions/450365}{450365}). Configuring various commands to produce color can be similarly difficult with each having its own ad hoc method to enable the command \enquote{to display colored output} (\href{https://unix.stackexchange.com/questions/94299}{94299}). Further complications are caused by using commands in combination. Commands like \verb|less| need to be configured to \enquote{properly display colours} (\href{https://unix.stackexchange.com/questions/150328}{150328}) (see also Section~\ref{contextual-behavior}). A consideration that is equally important in non-terminal contexts, there is a desire to choose colors that are not \enquote{impossible to read} (\href{https://unix.stackexchange.com/questions/94498}{94498}) in general as well as for those with \enquote{red green color blindness} (\href{https://unix.stackexchange.com/questions/3692}{3692}), for example. Choosing appropriate colors may involve detecting other colors that are already being used (\href{https://unix.stackexchange.com/questions/1755}{1755}, \href{https://unix.stackexchange.com/questions/172548}{172548}). There are cases where \enquote{removing control chars (including console codes / colours)} (\href{https://unix.stackexchange.com/questions/14684}{14684}) or converting \enquote{colored output of [an] arbitrary program} (\href{https://unix.stackexchange.com/questions/44956}{44956}) to other representations is required.


\subsubsection{\textbf{graphics (posts: 9, views: 1,641,946)}}

Given that terminal emulators are graphical applications capable of displaying color a natural next question is, \enquote{Are they capable of displaying graphics?} Some video display terminals had the ability to display graphics. In addition to character graphics~\cite{Wikipedia:Semigraphics}, two techniques that survive in some terminal emulators are the Sixel bitmap graphics format~\cite{Wikipedia:Sixel} and the ReGIS vector graphics markup language~\cite{Wikipedia:ReGIS}. Some terminal emulators have added further extensions~\cite{KittyGraphics}. Intermingling textual and graphical elements adds additional complexity which may be why many terminal emulators choose to avoid supporting graphics. Still, the desire remains to \enquote{view images from the terminal} (\href{https://unix.stackexchange.com/questions/35333}{35333}), \enquote{watch movies inside the terminal} (\href{https://unix.stackexchange.com/questions/123674}{123674}), and \enquote{draw images, shapes etc in a terminal} (\href{https://unix.stackexchange.com/questions/146486}{146486}).


\subsubsection{\textbf{window-size (posts: 21, views: 1,073,338)}}

Windowed terminals can have as many rows and columns as will fit on screen or be shrunk down to barely usable dimensions. Some commands need to be configured to use \enquote{the full width of the terminal} (\href{https://unix.stackexchange.com/questions/9301}{9301}). In some situations, users may wish to \enquote{set the maximum line width} (\href{https://unix.stackexchange.com/questions/222528}{222528}) in order to force a command to ignore the full size of the terminal. When lines are longer than the terminal is wide, should they be wrapped (\href{https://unix.stackexchange.com/questions/20493}{20493}, \href{https://unix.stackexchange.com/questions/70419}{70419})? Some terminal-based applications perform their own \enquote{long line wrapping} (\href{https://unix.stackexchange.com/questions/122795}{12279$^*$}). Terminal windows may be resized while running with a corresponding expectation that the content will be dynamically reformatted \enquote{on terminal dimension changes} (\href{https://unix.stackexchange.com/questions/11389}{11389}).


\subsubsection{\textbf{fonts (posts: 16, views: 444,840)}}

From a terminal perspective, encodings like UTF-8 (see Section~\ref{encodings}) solve one half of a problem, defining which byte sequences correspond to which characters. The other half of the problem is correctly displaying the glyphs corresponding to those byte sequences. Many of these issues are terminal emulator specific. More universal issues include knowing what font the terminal is using (\href{https://unix.stackexchange.com/questions/96962}{9696$^*$}, \href{https://unix.stackexchange.com/questions/154736}{154736}), knowing how to change the terminal font (\href{https://unix.stackexchange.com/questions/372447}{372447$^*$}, \href{https://unix.stackexchange.com/questions/407710}{407710}), finding \enquote{the best font for rendering a codepoint} (\href{https://unix.stackexchange.com/questions/162305}{162305}), problems with non-monospace fonts (\href{https://unix.stackexchange.com/questions/273059}{273059}), and enabling or disabling more advanced font rendering features, such as ligatures (\href{https://unix.stackexchange.com/questions/278175}{278175}).


\subsubsection{\textbf{mouse (posts: 11, views: 330,779)}}

With graphical environments comes the mouse (or similar pointing device) and the expectation that the mouse should be able to interact with the text in the terminal window. Unlike other applications, many terminal-based applications do not allow \enquote{placing the cursor by clicking} (\href{https://unix.stackexchange.com/questions/444601}{444601}) on a position with the mouse. Some terminal-based applications do support the mouse but that support may need to be enabled (\href{https://unix.stackexchange.com/questions/252995}{252995}) or disabled by users who prefer that the terminal handle \enquote{mouse-select and copy-paste} (\href{https://unix.stackexchange.com/questions/187695}{187695}). When the terminal is handling the mouse, users may wish to configure the text that the terminal selects in response to mouse activity (\href{https://unix.stackexchange.com/questions/45925}{45925}, \href{https://unix.stackexchange.com/questions/91137}{91137}, \href{https://unix.stackexchange.com/questions/290544}{290544}) which may be complicated by the presence of other applications (\href{https://unix.stackexchange.com/questions/318281}{318281}).


\subsubsection{\textbf{cut-copy-paste (posts: 67, views: 1,612,489)}}

Terminal-based applications implementing copy\&paste often do not use the clipboard. Due to independent implementations, copy\&paste differs between terminal-based applications, falling back to the terminal to provide it (\href{https://unix.stackexchange.com/questions/163715}{163715}).

To further complicate matters, the X Window System not only has a clipboard but a primary \emph{and} secondary selection~\cite{TheSecondarySelection} with Wayland also providing support for a primary selection in addition to the clipboard~\cite{WaylandMisconceptionsDebunked}. There is an obvious desire to \enquote{use just one unified clipboard} (\href{https://unix.stackexchange.com/questions/13585}{13585}) and have the \enquote{terminal selection automatically copy to the system clipboard} (\href{https://unix.stackexchange.com/questions/172044}{172044}) or, at a minimum, simply for consistent behavior from application to application (\href{https://unix.stackexchange.com/questions/178070}{178070}). These complications in addition to the peculiarities of some terminal-based applications create a situation where copying and pasting between the terminal and other applications \enquote{does not always work} (\href{https://unix.stackexchange.com/questions/136229}{136229}) as expected. Copying between the terminal and other applications also raises conversion questions (\href{https://unix.stackexchange.com/questions/293472}{293472}).

Specific to the terminal, it is not always easy or obvious how to \enquote{copy multi-page text from the terminal} (\href{https://unix.stackexchange.com/questions/109935}{109935}) or how to copy the text faithfully and not \enquote{as displayed} with characters that are a result of padding, line wrapping, or other display issues (\href{https://unix.stackexchange.com/questions/126577}{126577}, \href{https://unix.stackexchange.com/questions/218248}{218248}). There is a desire to access the system clipboard from the terminal (\href{https://unix.stackexchange.com/questions/125839}{125839}) and easily \enquote{copy a file or directory location as a path and then paste it in a terminal window} (\href{https://unix.stackexchange.com/questions/102551}{102551}). When copying relative file paths from the terminal the working directory of the terminal-based program that produced the relative file path is needed \enquote{to construct a full path} (\href{https://unix.stackexchange.com/questions/211418}{211418}) (see also Section \ref{working-directory}). There is a desire to quickly copy the last command from the terminal (\href{https://unix.stackexchange.com/questions/91863}{91863}) and paste commands into the terminal and either have them automatically executed (\href{https://unix.stackexchange.com/questions/364047}{364047}, \href{https://unix.stackexchange.com/questions/633370}{633370}, \href{https://unix.stackexchange.com/questions/689394}{689394$^*$}) or prevent their automatic execution (\href{https://unix.stackexchange.com/questions/96357}{96357}, \href{https://unix.stackexchange.com/questions/202732}{202732}, \href{https://unix.stackexchange.com/questions/230678}{230678}, \href{https://unix.stackexchange.com/questions/309786}{309786$^*$}). Bracketed paste was developed to address the problem of differentiating between typed and pasted text~\cite{Wikipedia:BracketedPaste} and is what now prevents the automatic execution of pasted commands. While this distinction is important in some situations (\href{https://unix.stackexchange.com/questions/355610}{355610}) bracketed paste can cause unexpected behavior in terminal-based applications that do not support it (\href{https://unix.stackexchange.com/questions/196098}{196098}). When pasting text as input to a terminal-based application the pasted text \enquote{gets mixed with the final output} (\href{https://unix.stackexchange.com/questions/693720}{693720}) causing the two to be confusingly intermingled. An \enquote{ancient Unix limitation} also limits lines to 4,096 characters (\href{https://unix.stackexchange.com/questions/131105}{131105}, \href{https://unix.stackexchange.com/questions/643777}{643777}). Given the heavy use of the keyboard some wish to \enquote{select text [...] via the keyboard only} (\href{https://unix.stackexchange.com/questions/210768}{210768}) and perform other operations such as switching \enquote{back and forth between [...] tabs using the keyboard} (\href{https://unix.stackexchange.com/questions/67861}{67861}\footnote{key-binding question\label{key-binding-footnote}}).

The terminal's keyboard shortcuts differ from those in most other applications. This is perhaps not surprising when we consider that the terminal predates many of these other applications (\href{https://unix.stackexchange.com/questions/16062}{16062}). A frequent stumbling block, Ctrl-C, the keyboard shortcut used to copy in most applications, is the terminal's default shortcut for interrupting a command. Ctrl-V, which users might expect to paste a previously cut or copied selection, is used by popular shells like \verb|bash| to mean \enquote{insert the next character verbatim}. The serendipitous existence and use of the Command key on the Macintosh allows it to circumvent some of these issues.

However, the issue remains for the vast majority of users. Often there is a desire simply for consistency (\href{https://unix.stackexchange.com/questions/52673}{52673$^*$}, \href{https://unix.stackexchange.com/questions/271150}{271150}, \href{https://unix.stackexchange.com/questions/339160}{339160}) and a \enquote{[s]ingle set of keyboard shortcuts for copy/paste/cut across all [...] applications} (\href{https://unix.stackexchange.com/questions/33501}{33501}). Previous attempts to address these issues have not been successful (\href{https://unix.stackexchange.com/questions/18589}{18589}\footref{key-binding-footnote}).


\subsubsection{\textbf{key-bindings (posts: 30, views: 677,455)}}

Not limited to shortcuts for copy and paste, at one point even the default behavior of the backspace and delete keys when using the terminal was potentially problematic~\cite{Archive:ConsistentBackSpaceandDeleteConfiguration}. Some keys, like the meta key, that were present on keyboards of the day but are no longer typical, have to be remapped to other keys, like the alt or escape key (\href{https://unix.stackexchange.com/questions/28993}{28993}). Different terminals send different byte sequences for the home and end keys (\href{https://unix.stackexchange.com/questions/20298}{20298}), as well as other special keys (\href{https://unix.stackexchange.com/questions/198711}{198711}). In addition, not all key \emph{combinations} have a standard representation (\href{https://unix.stackexchange.com/questions/79374}{79374}) (although that situation appears to be improving) while others overlap (\href{https://unix.stackexchange.com/questions/226327}{226327}). Different keyboard layouts change what is produced when a physical key is pressed. Opinions differ on whether shortcuts should be bound to the physical key (\href{https://unix.stackexchange.com/questions/91355}{91355}, \href{https://unix.stackexchange.com/questions/214273}{214273}).

\subsection{The Terminal as Dutiful Scribe}

There are many \enquote{full-screen} terminal applications but interactions with the terminal are often much less complicated. The terminal is frequently used in a command and response scenario that some refer to as typescript-style interaction~\cite{Acme}. In this scenario, the role of the terminal is that of a dutiful scribe --- preserving a record of commands issued and their output.


\subsubsection{\textbf{scrollback (posts: 73, views: 3,304,743)}} \label{scrollback}

Scrollback, a record of commands issued and their output along with previous prompts, can become quite cluttered (\href{https://unix.stackexchange.com/questions/107080}{107080}) with users looking for techniques to quickly \enquote{scroll to the last command} (\href{https://unix.stackexchange.com/questions/287541}{287541$^*$}), search scrollback (\href{https://unix.stackexchange.com/questions/12848}{12848}), save \enquote{terminal output to a file} (\href{https://unix.stackexchange.com/questions/200637}{200637}), distinguish normal output from errors (\href{https://unix.stackexchange.com/questions/290696}{290696}), and provide additional context such as \enquote{exit status code after each command} (\href{https://unix.stackexchange.com/questions/703918}{703918}). Sensitive information may be used when issuing a command or appear as part of a command's output. The question then becomes how to properly purge this record (\href{https://unix.stackexchange.com/questions/26975}{26975}).

When it is known that a command will produce many \emph{pages} of output a \emph{pager} can be used. Ideally a pager would only be invoked when required (\href{https://unix.stackexchange.com/questions/245064}{245064}) --- a similar effect can be achieved by configuring a pager to quit when it determines it is not required (\href{https://unix.stackexchange.com/questions/107315}{107315}). Some commands like \verb|git| and \verb|man| will pipe their output to a pager by default but there are those who wish they did not (\href{https://unix.stackexchange.com/questions/23394}{23394}). Pagers also come with their own set of issues (\href{https://unix.stackexchange.com/questions/23304}{23304}, \href{https://unix.stackexchange.com/questions/96635}{96635}, \href{https://unix.stackexchange.com/questions/104499}{104499}, \href{https://unix.stackexchange.com/questions/134385}{134385}, \href{https://unix.stackexchange.com/questions/158583}{158583}, \href{https://unix.stackexchange.com/questions/169886}{169886}, \href{https://unix.stackexchange.com/questions/197199}{197199$^*$}, \href{https://unix.stackexchange.com/questions/257788}{257788}, \href{https://unix.stackexchange.com/questions/412060}{412060}, \href{https://unix.stackexchange.com/questions/429466}{429466}). 

Pagers and other \enquote{full-screen} terminal applications must implement and manage their own scrollback if required (\href{https://unix.stackexchange.com/questions/145181}{145181}). These applications often switch to the terminal's \enquote{alternate screen} which has no scrollback. If the alternate screen is disabled or not used, the most recent screen contents for a full-screen application will become part of the scrollback buffer when the application terminates~\cite{XTermFAQAlternateScreen}.

In the more typical scenario, the screen appears to clear when a full-screen application switches away from the alternate screen as it terminates (\href{https://unix.stackexchange.com/questions/27941}{27941}, \href{https://unix.stackexchange.com/questions/60533}{60533}). While enough seem to prefer the screen \enquote{clearing} behavior (\href{https://unix.stackexchange.com/questions/60499}{60499}) that it has largely become the default there are situations where restoring the \enquote{previous screen upon exit} is not desired (\href{https://unix.stackexchange.com/questions/59203}{59203}). Use of the alternate screen can be inconsistent across applications and systems resulting in the need to \enquote{configure the restore behavior consistently} (\href{https://unix.stackexchange.com/questions/85398}{85398}). Accidentally switching to the alternate screen may result in users wondering why \enquote{[s]crolling is disabled all of a sudden} (\href{https://unix.stackexchange.com/questions/259922}{259922}).

In a terminal, previous lines scroll up, mimicking paper on a teletype. As a result, the terminal prompt has a tendency to migrate to the bottom of its window. There are some who wish this orientation could be reversed \enquote{like a normal terminal only upside down} (\href{https://unix.stackexchange.com/questions/324878}{324878}) with the prompt \enquote{at the top of the terminal} (\href{https://unix.stackexchange.com/questions/307177}{307177$^*$}). Even though terminals mostly deal with \enquote{just text}, the volume and the speed at which that text can be produced mean that even modern terminal emulators have performance issues to consider (\href{https://unix.stackexchange.com/questions/41225}{41225}) with many benchmarks focusing on scrolling speed~\cite{WarpPerformance}. One issue that remains unsolved is how to easily use \enquote{text from previous commands' output} (\href{https://unix.stackexchange.com/questions/385}{385}) with, or when constructing a new command (see the division of labor discussed in Section~\ref{command-history}).


a) \textbf{flow-control (posts: 7, views: 1,015,024)}
An older solution, called flow control~\cite{TheTTYDemystified}, allows users to \enquote{pause the output} (\href{https://unix.stackexchange.com/questions/294625}{294625}) of the terminal. Useful in some cases, this feature is the cause of posts asking how to \enquote{unfreeze [the terminal] after accidentally pressing Ctrl-S} (\href{https://unix.stackexchange.com/questions/12107}{12107}), wondering why the feature exists at all (\href{https://unix.stackexchange.com/questions/137842}{137842}), and asking how to \enquote{permanently disable} it (\href{https://unix.stackexchange.com/questions/332791}{332791}).


\subsubsection{\textbf{command-history (posts: 35, views: 2,533,145)}} \label{command-history}

The division of labor that has evolved between the terminal and the shell is most likely not what would be designed from scratch. The terminal's canonical mode provides simple line editing but shells and other command-line interfaces often implement their own, more advanced, command history, line editing, and completion. A common desire is to \enquote{maintain the same history across multiple terminals} (\href{https://unix.stackexchange.com/questions/131504}{131504}) but opinions differ on when the synchronization should occur (\href{https://unix.stackexchange.com/questions/212646}{212646}). There are users who prefer to not share history across terminals (\href{https://unix.stackexchange.com/questions/295165}{295165}) and some who would prefer a \enquote{[p]er-directory history} (\href{https://unix.stackexchange.com/questions/71189}{71189}). Also common is the desire to save more than the default number of commands (\href{https://unix.stackexchange.com/questions/20861}{20861}). With command history being a rolling buffer there is sometimes a desire to \enquote{save the last command} (\href{https://unix.stackexchange.com/questions/38072}{38072$^*$}) to a separate, more permanent location. Searching command history is possible but how \enquote{to cycle through} (\href{https://unix.stackexchange.com/questions/73498}{73498}) results or continue searching (\href{https://unix.stackexchange.com/questions/47814}{47814}) is not intuitive for some users, and we can see others looking for alternate behavior, such as searching \enquote{for a previous command with [what was] just typed} (\href{https://unix.stackexchange.com/questions/231605}{231605}). Different shells (as well as other command-line interfaces) have different default command history behavior, and customizing command history behavior can be difficult because of the amount of shell knowledge  (\href{https://unix.stackexchange.com/questions/265957}{265957}) and in some case line editing library knowledge (\href{https://unix.stackexchange.com/questions/420356}{420356}) required.

Sensitive information may be used when issuing a command. A natural question is how to \enquote{temporarily suspend history tracking} (\href{https://unix.stackexchange.com/questions/10922}{10922}). The way this is typically implemented can cause confusion when encountered by accident with users wondering why \enquote{commands that start with spaces} (\href{https://unix.stackexchange.com/questions/115917}{115917}) are not being added to the command history. Similarly, users need a way to \enquote{clear the history for the current session} (\href{https://unix.stackexchange.com/questions/544373}{544373}) or \enquote{close a terminal without saving the history} (\href{https://unix.stackexchange.com/questions/25049}{25049}). Some commands are more dangerous than others, and simply recording them in command history makes it possible to inadvertently run them again (\href{https://unix.stackexchange.com/questions/92742}{92742$^*$}). There are also those who would like a way to \enquote{save [a] command for later use} without executing it (\href{https://unix.stackexchange.com/questions/569941}{569941$^*$}).


\subsubsection{\textbf{prompt (posts: 24, views: 1,749,193)}}

Customizing the shell prompt is a gateway to many terminal issues (\href{https://unix.stackexchange.com/questions/31695}{31695}), through the peculiarities of the Unix shell (\href{https://unix.stackexchange.com/questions/346924}{346924}), and sometimes of a particular shell such as how \verb|bash| (\href{https://unix.stackexchange.com/questions/105926}{105926}) or \verb|zsh| (\href{https://unix.stackexchange.com/questions/90772}{90772$^*$}) calculate the prompt's display length. Some users are frustrated by the prompt's tendency to move around and wish that the prompt was always \enquote{at the bottom} (\href{https://unix.stackexchange.com/questions/153102}{153102}) or \enquote {at the top} (\href{https://unix.stackexchange.com/questions/218323}{218323}) of the terminal (see also Section~\ref{scrollback}). The line-oriented nature of the terminal and command-oriented nature of the shell can cause confusion with users wondering why they are being \enquote{repeatedly prompted with \textgreater} (\href{https://unix.stackexchange.com/questions/158997}{158997}). When the prompt is constructed dynamically, there are performance issues to consider (\href{https://unix.stackexchange.com/questions/565905}{565905}).


\subsubsection{\textbf{line-editing (posts: 31, views: 529,774)}}

With the customization possible, it is interesting that most use the \verb|emacs| or \verb|vi| line-editing keybindings. Sometimes this is by default. Even when the possibilities are known (\href{https://unix.stackexchange.com/questions/85390}{85390}) there is the further step of knowing how to apply the required configuration (\href{https://unix.stackexchange.com/questions/43003}{43003}). These keybindings are internally consistent and have a long history but are inconsistent with most other, non-terminal, applications.

Line-editing libraries~\cite{GNUReadline,Github:Editline,Github:Linenoise,Github:Liner} allow for consistency across terminal applications  (\href{https://unix.stackexchange.com/questions/22740}{22740}). However, customizations are then specific to each line-editing library. There are inevitable differences (\href{https://unix.stackexchange.com/questions/252546}{252546}) and a just as inevitable desire for \enquote{the same behavior} (\href{https://unix.stackexchange.com/questions/264791}{264791}) or at the very least the same functionality (\href{https://unix.stackexchange.com/questions/42629}{42629}). Not helping the situation is a division of labor that sees some sequences handled by the terminal and others by the terminal application leaving users wondering why \enquote{the arrow keys don't work} (\href{https://unix.stackexchange.com/questions/21733}{21733}) or how to \enquote{erase a mistyped invisible password} (\href{https://unix.stackexchange.com/questions/50493}{50493$^*$}). The command-oriented nature of the terminal makes it difficult to execute commands with fewer than two keystrokes (\href{https://unix.stackexchange.com/questions/89622}{89622}, \href{https://unix.stackexchange.com/questions/672295}{672295}) particularly when substantially different applications may use the same line-editing library.


\subsubsection{\textbf{completion (posts: 17, views: 326,981)}}

One of the most important features of modern line-editing libraries is completion, or tab-completion, as it is often called, after the key typically used to invoke it. Under the hood, there is a fair bit of machinery that makes completion possible which makes failures more difficult to debug (\href{https://unix.stackexchange.com/questions/499195}{499195}). Again, different applications and libraries implement this feature in different ways (\href{https://unix.stackexchange.com/questions/24419}{24419$^*$}).

Interacting with the completion machinery can be difficult. For example, cancelling \enquote{completion, but only completion} (\href{https://unix.stackexchange.com/questions/91161}{91161}), knowing how to \enquote{redirect command completion output} (\href{https://unix.stackexchange.com/questions/127506}{127506$^*$}), or how to \enquote{create a completion script} for a custom command (\href{https://unix.stackexchange.com/questions/136466}{136466$^*$}). The division of labor that results in completion being implemented by terminal applications rather than the terminal itself also means that the completion machinery cannot (easily) access the contents of the scrollback buffer (\href{https://unix.stackexchange.com/questions/29878}{29878}, \href{https://unix.stackexchange.com/questions/41886}{41886}).

\subsection{Context, Environment, and the Working Directory}


\subsubsection{\textbf{contextual-behavior (posts: 26, views: 589,235)}} \label{contextual-behavior}

According to the Unix philosophy, we should expect \enquote{the output of every program to become the input to another, as yet unknown, program [and not] clutter output with extraneous information}~\cite{UnixTimeSharingSystemForward}.
In practice, a program may want to provide some \enquote{extraneous information} to make the output more human-readable (as well as potentially performing other operations) when it knows that it \enquote{is `connected' to a terminal} (\href{https://unix.stackexchange.com/questions/401934}{401934}). This modal behavior adds complexity but is undeniably convenient and can even be used to stop common mistakes (\href{https://unix.stackexchange.com/questions/638221}{638221}) (see also Section~\ref{control-sequences}). It is also not without precedent --- Unix has other short-circuits (\href{https://unix.stackexchange.com/questions/29964}{29964}) and contextual behavior baked in (\href{https://unix.stackexchange.com/questions/515778}{515778}). When a program defaults to the wrong mode users may wish to \enquote{trick a command into thinking its output is going to a terminal} (\href{https://unix.stackexchange.com/questions/249723}{249723}) or vice-versa (\href{https://unix.stackexchange.com/questions/68574}{68574}). Some commands side-step this by sending information more likely to be consumed by a human to the stream more likely to be connected to a terminal (\href{https://unix.stackexchange.com/questions/166359}{166359}). In addition to detecting the presence of a terminal, a program may wish to know the specific type of terminal (\href{https://unix.stackexchange.com/questions/47037}{47037}).


\subsubsection{\textbf{environment-variables (posts: 36, views: 1,607,188)}}

The \verb|TERM| environment variable can be inspected but, much like the HTTP User-Agent header, it can be set to anything and so cannot be completely trusted (\href{https://unix.stackexchange.com/questions/23763}{23763}). The \verb|TERM| environment variable functions more as a hint to those terminal-based programs that choose to inspect it (\href{https://unix.stackexchange.com/a/43951}{43951}). Other environment variables play important roles. The \verb|EDITOR| and \verb|VISUAL| environment variables allow users to \enquote {set the default [terminal-based] editor} (\href{https://unix.stackexchange.com/questions/501862}{501862}) and the \verb|PATH| environment variable determines which commands can be invoked easily (\href{https://unix.stackexchange.com/questions/45578}{45578}). Many users have questions about the best \enquote{way to set environment variables} (\href{https://unix.stackexchange.com/questions/88201}{88201}) as well as where and how the initial values for environment variables are determined (\href{https://unix.stackexchange.com/questions/198794}{198794}, \href{https://unix.stackexchange.com/questions/246751}{246751}, \href{https://unix.stackexchange.com/questions/277944}{277944$^*$}). Approaches like \verb|path_helper| (\href{https://unix.stackexchange.com/questions/22979}{22979}) and \verb|environment.d| (\href{https://unix.stackexchange.com/questions/317282}{317282}) have the potential to alleviate some of this confusion but in the short term they add additional complexity. Given the importance of environment variables in the terminal environment there is an understandable desire to open a new terminal with an environment \enquote{that is an exact clone} (\href{https://unix.stackexchange.com/questions/6097}{6097}) of the current environment or to revert to a last known good environment by clearing \enquote{out all variables without closing terminal} (\href{https://unix.stackexchange.com/questions/172655}{172655}).


\subsubsection{\textbf{working-directory (posts: 16, views: 1,011,645)}} \label{working-directory}

Like all Unix processes, the terminal emulator process has a current working directory. 

However, when asking if there is a way to open a \enquote{new terminal window with the same directory as the previous window} (\href{https://unix.stackexchange.com/questions/327313}{327313}), open a terminal in \enquote{a directory other than \$HOME} (\href{https://unix.stackexchange.com/questions/140602}{140602}), change directories (\href{https://unix.stackexchange.com/questions/81224}{81224}), or be notified when entering a specific directory (\href{https://unix.stackexchange.com/questions/18532}{18532}) it is really the working directory of the application running in the terminal (usually a shell) that is being discussed.

\subsection{Control Sequences, UTF-8, and the Shell}


\subsubsection{\textbf{control-sequences (posts: 47, views: 1,269,176)}} \label{control-sequences}

Enabling much of the terminal's functionality are the in-band control sequences created for the DEC VT series of terminals~\cite{Wikipedia:VT100}. A subset of these were standardized~\cite{ECMA48} and are often referred to as ANSI escape codes or sequences~\cite{Wikipedia:ANSIEscapeSequences}. The xterm terminal emulator supports a \enquote{kind of amalgam of ANSI and the VT-whatever standards} (\href{https://unix.stackexchange.com/questions/5802}{5802}) plus extensions~\cite{XtermControlSequences, EscapeSequences}. Later terminal emulators followed xterm's lead.
As with useful features in other applications, control sequences can, unfortunately, be abused~\cite{TerminalEmulatorSecurityIssues} with a particular implementation exposing a vulnerability~\cite{DontTrustThisTitle}. There is an obvious desire to \enquote{avoid escape sequence attacks} (\href{https://unix.stackexchange.com/questions/15101}{15101}). Thankfully, as with other products, security issues are often fixed soon after being found, and work to find additional vulnerabilities continues~\cite{ANSITerminalSecurity}. While not as severe as security-related issues, binary data may contain unintended control sequences which, if accidentally sent to the terminal, can wreak havoc on terminal settings and leave users wondering how to fix the \enquote{terminal after displaying a binary file} (\href{https://unix.stackexchange.com/questions/79684}{79684$^*$}) or how to \enquote{prevent random console output from breaking the terminal} (\href{https://unix.stackexchange.com/questions/247999}{247999}). To avoid these problems some applications avoid outputting \enquote{binary} data unless explicitly instructed to do so (\href{https://unix.stackexchange.com/questions/217936}{217936}) (see also Section~\ref{contextual-behavior}). Even when the terminal is functioning as intended, misbehaving or misused applications may cause unexpected behavior (\href{https://unix.stackexchange.com/questions/164944}{164944}, \href{https://unix.stackexchange.com/questions/469770}{469770}).

As the \enquote{escape} in \enquote{escape code} or \enquote{escape sequence} implies the vast majority of control sequences start with an escape character. Unfortunately, it is also possible to generate the escape character by pressing the escape key which exists on most keyboards. To distinguish between the escape key being pressed and the start of a control sequence many applications delay processing the escape character (\href{https://unix.stackexchange.com/questions/318433}{318433}). While the control sequences supported by terminal emulators have converged to support popular terminal-based applications, a terminal-based application may wish to \enquote{programmatically check} (\href{https://unix.stackexchange.com/questions/228369}{228369}) \enquote{what features a terminal supports} (\href{https://unix.stackexchange.com/questions/390360}{390360}).


\subsubsection{\textbf{encodings (posts: 29, views: 615,415)}} \label{encodings}

Many see 2008 as the year when UTF-8 became the rule rather than the exception~\cite{Wikipedia:UTF-8}. This shift seems to have solved many previous terminal encoding issues. However, there are still some issues that remain. Among the remaining issues is the question of how to easily \enquote{type Unicode characters} (\href{https://unix.stackexchange.com/questions/12244}{12244}) beyond the limited set that are easily possible from the keyboard. A related issue is understanding the byte sequences that correspond to Unicode characters and how to represent these in other contexts (\href{https://unix.stackexchange.com/questions/530788}{530788$^*$}). Knowing \enquote{the display width of a string of characters} (\href{https://unix.stackexchange.com/questions/245013}{245013}) is not simply about counting bytes.

Not all languages are left to right (\href{https://unix.stackexchange.com/questions/100811}{100811}, \href{https://unix.stackexchange.com/questions/167856}{167856}). Even with UTF-8 as the default, there are issues with non-printing characters and legacy control codes inherited from ASCII that are \enquote{invisible in terminal output} (\href{https://unix.stackexchange.com/questions/128019}{128019}, \href{https://unix.stackexchange.com/questions/723082}{723082}) as well as files in legacy formats (\href{https://unix.stackexchange.com/questions/184538}{184538}). Although support for UTF-8 is now nearly universal there is an understandable desire to be able to detect a terminal's level of Unicode support (\href{https://unix.stackexchange.com/questions/184345}{184345}).


\subsubsection{\textbf{quoting (posts: 11, views: 475,872)}}

As the most common terminal application, the shell determines how characters are interpreted, which are best avoided (\href{https://unix.stackexchange.com/questions/148043}{148043}), and what is required to \enquote{use a special character} (\href{https://unix.stackexchange.com/questions/296141}{296141}). Misunderstanding how the shell operates can easily cause confusion (\href{https://unix.stackexchange.com/questions/284638}{284638}, \href{https://unix.stackexchange.com/questions/446193}{446193}). In an effort to avoid some of this confusion, some terminal-based programs have started producing shell-friendly output (\href{https://unix.stackexchange.com/questions/258679}{258679}) (see also Section~\ref{contextual-behavior}).

\section{Discussion}

Despite its deceptively simple appearance, the terminal is a large and complex application. Using the core Wayland window system code~\cite{GitLab:Wayland} as a point of reference, at just over 100,000 lines of code the venerable xterm~\cite{XTerm} is approximately 2.5x larger than the core Wayland window system code. Even \verb|foot|, a relatively new, \enquote{lightweight and minimalistic Wayland terminal emulator}~\cite{Codeberg:Foot} is larger than the core Wayland window system code itself. The terminal's size, complexity, and long history result in what can seem like a bewildering array of issues. We have grouped and arranged these issues in an attempt to make them more digestible. We also believe that several themes emerge.

With the creation of the windowed terminal emulator the environment of the terminal changed. We claim that the terminal, as windowed terminal emulator, and the applications it supports, have yet to fully \textbf{adapt} to this new environment. We see evidence for this in posts asking why the mouse cannot be used to position the cursor when editing a command or why the terminal's keyboard shortcuts for copy and paste are different from most other applications. The long history of the terminal includes the accumulation of quirks and the evolution of conventions that often differ from conventions established outside the terminal. While users can and do adapt, this raises the question of how much of this is inherent and how much is accidental and avoidable. We see further evidence in posts asking if there is some way to make file names in terminal output clickable by default, as well as in the many posts looking for ways to integrate the terminal with other applications or asking for ways to synchronize command history or environment variable changes across multiple terminals.

While the terminal has not yet fully adapted to its new graphical environment neither, we claim, has it managed to completely \textbf{untangle} itself from the video-display terminal era. Instead of a video-display terminal with one login shell, there are now often multiple terminal windows for the same user. The \verb|jobs| command (even though it is provided by the shell) is particularly revealing here, in that it only shows jobs for the current shell in the current terminal and not all jobs on the current system for the current user (\href{https://unix.stackexchange.com/questions/509147}{509147}).

The line-oriented nature of the terminal and the command-oriented nature of interpreters like the shell forces each of these applications to re-implement much of the same functionality. The division of labor that has evolved between the terminal and the shell also makes it more cumbersome than one might expect to use the output of previous commands when constructing new commands. In general, it seems strange that terminal pagers are still required and that terminal features for interacting with scrollback are so minimal.

Terminal scrollback and command history are areas that seem underdeveloped and we believe more could be done to \textbf{extend} these features. This area has seen some activity recently: an improvement pioneered by FinalTerm and now adopted by many terminals, the \verb|OSC 133| control sequences~\cite{iTerm2:ProprietaryEscapeCodes} allow command interpreters to mark the beginning of the prompt, the beginning of a command, the start of the command output, and the end of the command output. This enables terminals with support for this feature to quickly scroll to these marks. In addition to better navigation and searching of scrollback and command history, the terminal could also provide more context in scrollback as well as do more to distinguish regular output from errors. We believe the terminal should also do more to prevent errors caused by misbehaving applications or unintentional control sequences.

More terminal application-related than terminal-related, we see a number of opportunities to \textbf{simplify} the terminal experience. These range from promoting better defaults to investigating how to standardize feature detection to developing tools to translate configuration into the myriad ad hoc settings required by various terminal-based applications.

Based on aggregate views it is clear that integration with other applications, job control, terminal multiplexing, and accessing remote systems are important use cases. Attempts to improve the terminal ignore these at their peril. More than that, any attempt to improve the terminal must take great care to preserve the vast majority of the terminal's functionality. We know from Hyrum's Law that \enquote{all observable behaviors of [a] system will be depended on by somebody}~\cite{HyrumsLaw}. Terminal-based applications have had half a century to develop a dependence on the terminal's behavior (\href{https://unix.stackexchange.com/questions/105325}{105325}, \href{https://unix.stackexchange.com/questions/278884}{278884}) - both documented and undocumented.

Finally, we see that some terminal-related issues are a result of not knowing where to look for more information. We hope that this paper may play a small part in rectifying that. We would also like to draw attention to resources like Thomas Dickey's \emph{XTerm Control Sequences}~\cite{XtermControlSequences} and articles like Linus Åkesson's \enquote{The TTY demystified}~\cite{TheTTYDemystified}. While terminal-related documentation does exist we believe what makes understanding the terminal more difficult is how this documentation often has to be pieced together as part of a historical record and that what is missing is comprehensive documentation of the terminal as it now exists.

In Section~\ref{related-work} we mentioned that a goal many current efforts to improve the terminal seem to share is improving the social and collaborative aspects of the terminal. We see little evidence of that as a prominent pain point.

\section{Limitations}

This paper is based on questions posted on U\&LSE and so inherits the biases of that community. As discussed previously, there are other similar sources of information but we believe U\&LSE to be the most relevant for the discussion in this paper. Our approach could be applied to other datasets to investigate how much the choice of U\&LSE affects our results.

Our findings are based on a subset of questions from U\&LSE. It is possible that this subset omits relevant questions. Pragmatically, it is not possible to evaluate every question. To mitigate against this threat our goal was to make our selection and categorization rules as clear as possible. In addition, our selection and categorization process can be reproduced using the scripts in our replication package.

Using regular expressions we were able to categorize 91.0\% of the posts identified by our selection process without manual intervention. Where possible we based our patterns on existing tags or common words or phrases occurring in more than one post. In cases where this was not possible, due to tags lacking in discriminatory power or idiosyncratic phrasing, we based our patterns on unique and identifiable passages still recognizable as being related to the category in question. Additionally, we found that the patterns we developed to identify a subset of all unrelated posts were sufficiently accurate to be used as a pre-filter. We exploited this characteristic to simplify the patterns needed for all other categories. The goal throughout the process was patterns that are easily understandable and stable. We do not claim that the patterns used are an optimal or minimal set.


It is possible for a question to belong to more than one category or for a post to contain more than one question. We found these threats to be more theoretical than practical. Only 7.9\% of selected posts were assigned more than one category. In the vast majority of cases, there was a clearly dominant category with additional categories aiding rather than hindering understanding of the issue. On the rare occasion where a post contained more than one question (\href{https://unix.stackexchange.com/questions/94299}{94299}, for example) we did not find our analysis affected.

It is possible to disagree with the categories we have chosen as well as our interpretation of the results. While we cannot anticipate any and all objections we have made our replication package available to foster transparency and enable others to reproduce, examine, and experiment with our approach and our results.


\section{Conclusion}

At one point in time, it was a commonly held belief that the terminal would go away --- replaced by sleeker interfaces. There are some who continue to hold this belief. Exceeding the strength of this belief is the terminal's stubborn refusal to do so. We believe the terminal has persisted because it provides a uniquely productive interface for many tasks. We also believe, and believe that this paper demonstrates, that a requirements-based understanding of how the terminal should evolve is possible.

Terminal lucidity \enquote{is an unexpected return of consciousness, mental clarity or memory shortly before death}~\cite{Wikipedia:TerminalLucidity}. We are not referring here to the death of the terminal. Far from it. The terminal's popularity seems, if anything, to be increasing. We are referring to the death of prejudices and preconceived notions about the terminal that are increasingly hard to maintain. As these fall away, we hope it will be possible to see the true potential of the terminal clearly.

\subsection*{Acknowledgements}

The authors thank Thomas Dickey for reviewing an early draft and suggesting ways to improve readability. The authors also gratefully acknowledge the support of the Natural Sciences and Engineering Research Council of Canada (NSERC, RGPIN-2021-03809) and the Swiss National Science Foundation (SNSF, Project ``FORCE'', SNF Project Number 232141).

\section{Appendix} \label{appendix}

The terminal is often mentioned but not the focus in questions about popular terminal-based applications like \verb|ssh| (\href{https://unix.stackexchange.com/questions/48863}{48863}, \href{https://unix.stackexchange.com/questions/71481}{71481}, \href{https://unix.stackexchange.com/questions/85920}{85920}, \href{https://unix.stackexchange.com/questions/108679}{108679}, \href{https://unix.stackexchange.com/questions/207365}{207365}, \href{https://unix.stackexchange.com/questions/317491}{317491}, \href{https://unix.stackexchange.com/questions/321968}{321968}, \href{https://unix.stackexchange.com/questions/323925}{323925}, \href{https://unix.stackexchange.com/questions/352298}{352298}, \href{https://unix.stackexchange.com/questions/371901}{371901}, \href{https://unix.stackexchange.com/questions/758893}{758893})
or \verb|vim| (\href{https://unix.stackexchange.com/questions/11369}{11369}, \href{https://unix.stackexchange.com/questions/25965}{25965}, \href{https://unix.stackexchange.com/questions/43033}{43033}, \href{https://unix.stackexchange.com/questions/186166}{186166}, \href{https://unix.stackexchange.com/questions/345494}{345494})
or applications like \verb|emacs| which may run in a terminal (\href{https://unix.stackexchange.com/questions/8424}{8424}, \href{https://unix.stackexchange.com/questions/174413}{174413}).
There are also questions about the Linux (or FreeBSD, etc.) console (\href{https://unix.stackexchange.com/questions/16861}{16861}, \href{https://unix.stackexchange.com/questions/23363}{23363}, \href{https://unix.stackexchange.com/questions/45228}{45228}, \href{https://unix.stackexchange.com/questions/49779}{49779}, \href{https://unix.stackexchange.com/questions/485156}{485156}).
or windowed terminal-emulator adjacent applications like PuTTY (\href{https://unix.stackexchange.com/questions/20257}{20257}, \href{https://unix.stackexchange.com/questions/143684}{143684}) that are not relevant to the discussion in this paper. Sometimes the terminal is simply mentioned incidentally (\href{https://unix.stackexchange.com/questions/3022}{3022}, \href{https://unix.stackexchange.com/questions/22561}{22561}, \href{https://unix.stackexchange.com/questions/72657}{72657}, \href{https://unix.stackexchange.com/questions/150501}{150501}, \href{https://unix.stackexchange.com/questions/164262}{164262}, \href{https://unix.stackexchange.com/questions/162979}{162979}, \href{https://unix.stackexchange.com/questions/233459}{233459}, \href{https://unix.stackexchange.com/questions/322814}{322814}, \href{https://unix.stackexchange.com/questions/369097}{369097}, \href{https://unix.stackexchange.com/questions/439486}{439486}, \href{https://unix.stackexchange.com/questions/569993}{569993}) or it, or a terminal-based application, is part of the situation but has no bearing on the problem or solution (\href{https://unix.stackexchange.com/questions/14629}{14629}, 
\href{https://unix.stackexchange.com/questions/23360}{23360}, \href{https://unix.stackexchange.com/questions/25998}{25998}, \href{https://unix.stackexchange.com/questions/34729}{34729}, \href{https://unix.stackexchange.com/questions/37539}{37539}, \href{https://unix.stackexchange.com/questions/52055}{52055}, \href{https://unix.stackexchange.com/questions/146193}{146193}, \href{https://unix.stackexchange.com/questions/166936}{166936}).
The terminal is mentioned in many questions that are really Unix, scripting, or shell-related questions (\href{https://unix.stackexchange.com/questions/10698}{10698}, \href{https://unix.stackexchange.com/questions/40708}{40708}, \href{https://unix.stackexchange.com/questions/70615}{70615}, \href{https://unix.stackexchange.com/questions/71492}{71492}, \href{https://unix.stackexchange.com/questions/72962}{72962}, \href{https://unix.stackexchange.com/questions/113556}{113556}, \href{https://unix.stackexchange.com/questions/134734}{134734}, \href{https://unix.stackexchange.com/questions/151757}{151757}, \href{https://unix.stackexchange.com/questions/153357}{153357}, \href{https://unix.stackexchange.com/questions/153508}{153508}, \href{https://unix.stackexchange.com/questions/155838}{155838}, \href{https://unix.stackexchange.com/questions/156084}{156084}, \href{https://unix.stackexchange.com/questions/157689}{157689}, \href{https://unix.stackexchange.com/questions/166853}{166853}, \href{https://unix.stackexchange.com/questions/189254}{189254}, \href{https://unix.stackexchange.com/questions/199085}{199085}, \href{https://unix.stackexchange.com/questions/245746}{245746}, \href{https://unix.stackexchange.com/questions/251691}{251691}, \href{https://unix.stackexchange.com/questions/283586}{283586}, \href{https://unix.stackexchange.com/questions/297729}{297729}, \href{https://unix.stackexchange.com/questions/377618}{377618}, \href{https://unix.stackexchange.com/questions/382191}{382191}, \href{https://unix.stackexchange.com/questions/411688}{411688}, \href{https://unix.stackexchange.com/questions/494378}{494378}, \href{https://unix.stackexchange.com/questions/498685}{498685}, \href{https://unix.stackexchange.com/questions/599013}{599013}).

In addition to the above questions, manually categorized as unrelated, 13 questions were manually corrected and assigned to the following categories: control-sequences (\href{https://unix.stackexchange.com/questions/105325}{105325}), encodings (\href{https://unix.stackexchange.com/questions/723082}{723082}), environment-variables (\href{https://unix.stackexchange.com/questions/23763}{23763}), integration (\href{https://unix.stackexchange.com/questions/152310}{152310}, \href{https://unix.stackexchange.com/questions/234136}{234136}, \href{https://unix.stackexchange.com/questions/268451}{268451}, \href{https://unix.stackexchange.com/questions/316401}{316401}), key-bindings (\href{https://unix.stackexchange.com/questions/214273}{214273}, \href{https://unix.stackexchange.com/questions/226327}{226327}, \href{https://unix.stackexchange.com/questions/262129}{262129}), prompt (\href{https://unix.stackexchange.com/questions/105958}{262129}), remote (\href{https://unix.stackexchange.com/questions/3212}{3212}), and scrollback (\href{https://unix.stackexchange.com/questions/12439}{12439}).

\balance{}
\bibliographystyle{IEEEtran}
\bibliography{references}

\end{document}